\documentclass[12pt]{article}
\usepackage{amssymb,amsmath,epsfig}
\usepackage{epsfig}
\usepackage{epstopdf}
\allowdisplaybreaks
\begin{document}
\title{\bf Thermodynamics of Ricci-Gauss-Bonnet Dark Energy}
\author{Ayesha Iqbal$^{1}$
\thanks{ayeshausmann@yahoo.com} and Abdul Jawad$^2$ \thanks{jawadab181@yahoo.com;~~abduljawad@ciitlahore.edu.pk}\\
$^{1}$ Department of Mathematics, Govt. College University,\\
Faisalabad, Pakistan.\\
$^2$ Department of Mathematics, COMSATS Institute of\\ Information
Technology, Lahore-54000, Pakistan.}

\date{}
\maketitle

\begin{abstract}
We investigate the validity of generalized second law of
thermodynamics of a physical system comprising of newly proposed
dark energy model called Ricci Gauss-Bonnet and cold dark matter
enveloped by apparent horizon and event horizon in flat
Friedmann-Robertson-Walker (FRW) universe. For this purpose,
Bekenstein entropy, Renyi, logarithmic and power law entropic
corrections are used. It is found that this law exhibits the
validity on both apparent and event horizons except for the case of
logarithmic entropic correction at apparent horizon. Also, we check
the thermodynamical equilibrium condition for all cases of entropy
and found its vitality in all cases of entropy.
\end{abstract}

\section{Introduction}

The revelation of black holes thermodynamics motivated the physicist
to examine the thermodynamics of cosmological models in accelerated
expanding universe \cite{1}-\cite{3}. Bekenstein and Hawking
determined that the entropy of black hole is proportional to its
event horizon \cite{4,5} which leads to important law named as
generalized second law of thermodynamics (GSLT) for black hole
physics. This law can be defined as the entropy of black hole and
its exterior is always increasing. The primitive level of
thermodynamics properties of horizons are exhibited by considering
Einstein field equations as alternate of first law of thermodynamics
\cite{6,7}. Gibbons and Hawking developed the Bekenstein's idea for
cosmological system by exhibiting that the entropy of cosmological
event horizon is proportional to horizon area \cite{8}. They
represented the equality of apparent horizon and event horizon for
de Sitter universe. The validity of GSLT was deeply studied later
\cite{9}-\cite{11}. GSLT in cosmological scenario implies that the
rate of change of entropy of horizon along with that of fluid inside
it will always greater than or equal to zero. Its mathematical
expression is
\begin{equation}\label{1}
\frac{dS_{horizon}}{dt}+\frac{dS_{inside}}{dt}\geq0.
\end{equation}

In addition, the holographic dark energy (HDE) is an interesting
effort in exploring the nature of dark energy in the framework of
quantum gravity. This model is motivated from the fundamental
holographic principle, that arises from black hole thermodynamics
and string theory \cite{19}-\cite{22}. HDE fascinated a large amount
of research despite of some objections \cite{23,24}. The choice of
the length scale $L$ appearing in the holographic dark energy
density $\rho_{de}=3M_{pl}L^{-2}$ gives rise to different dark
energy models. One of the its crucial model is holographic Ricci
dark energy model which is developed by assuming IR length scale as
the average radius of Ricci scalar curvature, $R^{-1/2}$
\cite{25}-\cite{27}. Moreover, its modified form is also presented
and discussed widely \cite{28}-\cite{30}.

Further, Wang et al. \cite{12} observed that GSLT is verified at
apparent horizon but not at event horizon for a specific model of
dark energy. In case of new holographic dark energy, GSLT is valid
fully on apparent horizon but partially on event horizon of universe
\cite{13}. The breakdown of GSLT was argued in case of event horizon
enveloping the universe as compared to apparent horizon \cite{14}.
Setare \cite{15} has derived the constraints on deceleration
parameter in order to fulfill GSLT in case of non-flat universe
enveloped by event horizon. The GSLT of thermodynamics has also been
analyzed in case of Braneworld \cite{16,17} and generally Levelock
gravity \cite{18}.

Moreover, modified matter part of Einstein Hilbert action results
dynamical models such as cosmological constants, quintessence,
k-essence, Chaplygin gas and holographic dark energy (HDE) models
\cite{49N}-\cite{bamba}. Moreover, several modified theories of
gravity are $f(R)$, $f(T)$ \cite{st1}, $f(R,\mathcal{T})$
\cite{ma1}, $f(G)$ \cite{mj1}, $f(T,T_G)$ \cite{R38},
$f(T,\emph{T})$ \cite{R19,R57} (where $R$ is the curvature scalar,
$T$ denotes the torsion scalar, $\mathcal{T}$ is the trace of the
energy momentum tensor and $G$ is the invariant of Gauss-Bonnet
defined as $G=R^2-4R_{\mu\nu}R^{\mu\nu}+ R_{\mu\nu
\lambda\sigma}R^{\mu\nu\lambda\sigma}$). For clear review of DE
models and modified theories of gravity, see the reference
\cite{bamba}. Some authors \cite{RRS16}-\cite{JJJ} have also
discussed various DE models in different frameworks and found
interesting results.

Recently, Saridakis \cite{31} Ricci-Gauss-Bonnet holographic dark
energy in which Infrared cutoff is determined by both Ricci scalar
and the Gauss-Bonnet invariant. Such a construction has the
significant advantage that the Infrared cutoff, and consequently the
HDE density, does not depend on the future or the past evolution of
the universe, but only on its current features, and moreover it is
determined by invariants, whose role is fundamental in gravitational
theories. This model has IR cutoff form as $\frac{1}{L^{2}}=-\alpha
R+\beta \sqrt{|G|}$ where $\alpha$ and $\beta$ are model parameters.
In flat FRW geometry, the Ricci scalar ($R$) and the Gauss-Bonnet
invariant ($G$) are given as $R=-6(2H^{2}+\dot{H})$ and
$G=24H^{2}(H^{2}+\dot{H})$, respectively \cite{31}.

In the present work, we examine the validity of GSLT by assuming
various forms of entropy on apparent and event horizons. We have
also examined that wether each entropy attain maximum (thermodynamic
equilibrium) by satisfying the condition $\ddot{S}_{tot}<0$. The
plan of the paper is as follows: In sections \textbf{2} and
\textbf{3}, we have examined the validity of GSLT as well as thermal
equilibrium condition at apparent and event horizons, respectively.
The results are summarized in the last section.

\section{Generalized Second Law of Thermodynamics at Apparent Horizon}

According to GSLT, the entropy of horizon and entropy of matter
resources inside horizen does not decrease with respect to time.
Following Eq.(\ref{1}), we can write
\begin{equation}\label{2}
\dot{S_{tot}}=\dot{S_{h}}+\dot{S_{in}}\geq 0.
\end{equation}
Here $\dot{S_{h}}$ gives entropy of horizon and entropy of matter
inside horizon is represented by $\dot{S_{in}}$. Now considering
spatially flat FRW universe, the first Friedmann equation is
\begin{equation}\label{4}
H^{2}=\frac{\kappa^{2}}{3}(\rho_{eff}+P_{eff}).
\end{equation}
Here $\rho_{eff}$ and $P_{eff}$ are effective density and pressure,
respectively. We have made the following two assumptions (i) an
entropy is associated with the horizon in addition to the entropy of
the universe inside the horizon, (ii) according to the local
equilibrium hypothesis, there is no spontaneous exchange of energy
between the horizon and fluid inside. Moreover, Gibb's equation can
be written as
\begin{equation}\label{5}
TdS_{in}=P_{eff}dV+dE_{in}.
\end{equation}
Here $E_{in}=\rho_{eff}V$, $V=\frac{4\pi}{3}R_{h}^{2}$ and
$T=\frac{1}{2\pi R_{h}}$ which modified above equation as follows
\begin{equation}\label{7}
\dot{S_{in}}=8\pi^{2}R_{h}^{3}(\rho_{eff}+P_{eff})(\dot{R_{h}}-HR_{h}).
\end{equation}
For flat FRW universe, the Hubble horizon can be defined as
\begin{equation}\label{9}
R_{h}=\frac{1}{H}.
\end{equation}
By utilizing above horizon, $P_{eff}=p_{de}$ (for cold dark matter
$p_{m}=0$) and $\rho_{eff}=\rho_{d}+\rho_{m}$ in Eq.(\ref{7}), we
can get
\begin{equation}\label{10}
\dot{S_{in}}=8\pi^{2}R_{h}^{3}\bigg(\rho_{eff}+\rho_{d}\omega_{de}\bigg)(\dot{R_{h}}-1).
\end{equation}
From conservation equation, one can obtain
\begin{equation}\label{11}
\dot{\rho_{d}}=3H(1+\omega_{d})\rho_{d}~~~\Rightarrow~~~
\omega_{d}=-1-\frac{\dot{\rho_{d}}}{3H\rho_{d}}.
\end{equation}
Substituting the value of $\omega_{de}$ in Eq.(\ref{10}), we get
\begin{equation}\label{13}
\dot{S_{in}}=8\pi^{2}R_{h}^{3}\bigg(\rho_{m}-\frac{\dot{\rho_{d}}}{3H}\bigg)(\dot{R_{h}}-1).
\end{equation}
Moreover, Ricci-Gauss Bonnet dark energy can be defined as follows
\cite{31}
\begin{equation}\label{14}
\rho_{d}=3\bigg(6 \alpha(2H^{2}+\dot{H})+2\sqrt{3}\beta
H\sqrt{|H^{2}+\dot{H}|}\bigg).
\end{equation}
Here $\alpha$ and $\beta$ are the model parameters. Standard Ricci
dark energy can be obtained by substituting $\beta=0$ and $\alpha=0$
yields a pure Gauss-Bonnet HDE. The density parameters can be
introduced as
\begin{equation}\label{15}
\Omega_{m}=\frac{\rho_{m}}{3H^{2}}, \quad
\Omega_{d}=\frac{\rho_{d}}{3H^{2}}.
\end{equation}
According to first Friedman equation, we can obtain
\begin{equation}\label{18}
\Omega_{d}+\Omega_{m}=1.
\end{equation}
Also, $\rho_{m}$ can be evaluated by using conservation equation as
follows
\begin{equation}\label{20}
\rho_{m}=\frac{\rho_{m_{0}}}{a^{3}},
\end{equation}
with $\rho_{m_{0}}=3H^2_{0}\Omega_{m_{0}}$. By using this value of
$\rho_{m}$, $\Omega_{m}$ takes the following form
\begin{equation}\label{21}
\Omega_{m}=\frac{\Omega_{m_{0}}H_{0}^{2}}{a^{3}H^{2}}.
\end{equation}
Using Eqs.(\ref{18}) and (\ref{21}), we can find $H$ as
\begin{equation}\label{22}
H=\frac{H_{0}\sqrt{\Omega_{m_{0}}}}{a^{3}(1-\Omega_{d})}.
\end{equation}
Differentiating $H$, we obtain
\begin{equation}\label{23}
\dot{H}=-\frac{H^{2}}{2}\bigg(3-\frac{\acute{\Omega_{d}}}{1-\Omega_{d}}\bigg),
\end{equation}
where prime denotes the differentiation with respect to $x=lna$.
Also, differentiation of $R_A$ with respect to $t$ leads to
\begin{equation}\label{24}
\dot{R_{h}}=-\frac{\dot{H}}{H^{2}}=\frac{1}{2}\bigg(3-\frac{\acute{\Omega_{d}}}{1-\Omega_{d}}\bigg).
\end{equation}
We get the following value of $\acute{\Omega_{de}}$ by
differentiating Eq.(\ref{15})
\begin{equation}\label{26}
\acute{\Omega_{d}}=\bigg(\frac{\dot{\rho_{d}}}{3H^{3}}
+\frac{\rho_{d}}{H^{2}}\bigg)\bigg(1+\frac{\rho_{d}}{3H^{2}(1-\Omega_{d})}\bigg)^{-1}.
\end{equation}
Now, $\dot{R_{h}}$ takes the form
\begin{equation}\label{27}
\dot{R_{h}}=\frac{1}{2}\bigg(3-\bigg(\frac{1}{1-\Omega_{d}}\bigg)
\bigg(\frac{\dot{\rho_{d}}}{3H^{3}}+\frac{\rho_{d}}{H^{2}}\bigg)
\bigg(1+\frac{\rho_{d}}{3H^{2}(1-\Omega_{d})}\bigg)^{-1}\bigg).
\end{equation}
Also, Friedman first equation gives $\rho_{m}=3H^{2}-\rho_{d}$ and
hence we can write
\begin{equation}\label{29}
\dot{S_{in}}=8(\pi)^{2}R_{h}^{3}\bigg(3H^{2}-\rho_{d}
-\frac{\dot{\rho_{d}}}{3H}\bigg)(\dot{R_{h}}-1).
\end{equation}
By inserting Eq.(\ref{9}) in above equation, we have
\begin{equation}\label{30}
\dot{S_{in}}=\frac{8(\pi)^{2}}{H^{3}}\bigg(3H^{2}-\rho_{d}-\frac{\dot{\rho_{d}}}{3H}\bigg)(\dot{R_{h}}-1).
\end{equation}
By using value of $\dot{R_{h}}$ from Eq.(\ref{27}), we get
\begin{eqnarray}\nonumber
\dot{S_{in}}&=&\frac{8\pi^{2}}{H^{3}}\bigg(3H^{2}-\rho_{d}-\frac{\dot{\rho_{d}}}{3H}\bigg)
\bigg(\frac{1}{2}-\bigg(\frac{1}{2(1-\Omega_{d})}\bigg)\bigg(\frac{\dot{\rho_{d}}}{3H^{3}}
+\frac{\rho_{d}}{H^{2}}\bigg)\\\label{31}
&\times&\bigg(1+\frac{\rho_{d}}{3H^{2}(1-\Omega_{d})}\bigg)^{-1}\bigg).
\end{eqnarray}
Next, we will discuss the various expressions of entropy-area
relations in order analyze the validity of GSLT on Hubble horizon.

\subsection{Bekenstein Entropy}

The Bekenstein entropy is given by
\begin{equation}\label{32}
S_{h}=\frac{A}{4G}.
\end{equation}
By using $G=1,~c=1$ and $A=4\pi R_{h}^{2}$ being the area of
horizon, we get
\begin{equation}\label{33}
S_{h}=\pi R^{2}_{h} ~~~~\Rightarrow~~~~ \dot{S}_{h}=2\pi
R_{h}\dot{R_{h}}.
\end{equation}
By using the expressions of $R_{h}$ and $\dot{R_{h}}$, we have
\begin{equation}\label{35}
\dot{S}_{h}=\frac{\pi}{H}
\bigg(3-\bigg(\frac{1}{1-\Omega_{d}}\bigg)\bigg(\frac{\dot{\rho_{d}}}{3H^{3}}+\frac{\rho_{d}}{H^{2}}\bigg)
\bigg(1+\frac{\rho_{d}}{3H^{2}(1-\Omega_{d})}\bigg)^{-1}\bigg).
\end{equation}
Equations (\ref{31}) and (\ref{35}) join to form
\begin{eqnarray}\nonumber
\dot{S_{tot}}&=&\frac{8\pi^{2}}{H^{3}}
\bigg(\frac{1}{2}-\bigg(\frac{1}{2(1-\Omega_{de})}\bigg)\bigg(3-\frac{\rho_{de}}{H^{2}}\bigg)
\bigg(1+\frac{\rho_{de}}{3H^{2}(1-\Omega_{de})}\bigg)^{-1}\bigg)\\\nonumber
&\times&\bigg(6H^{2}-\rho_{de}\bigg)+\frac{\pi}{H}
\bigg(3-\bigg(\frac{1}{1-\Omega_{de}}\bigg)
\bigg(1+\frac{\rho_{de}}{3H^{2}(1-\Omega_{de})}\bigg)^{-1}
\\\label{36}
&\times&\bigg(3-\frac{\rho_{de}}{H^{2}}\bigg)\bigg).
\end{eqnarray}
where $\dot{S}_{tot}$ represents the total entropy, i.e.,
$\dot{S}_{tot}=\dot{S}_{in} + \dot{S}_h$.

Now, we assume the power law form of scale factor, i.e.,
$a=a_{0}t^{n}$, where $n$ and $a_{0}$ appear as constant parameters.
Under this assumption, the values of $H$ and $R_{h}$ turns out to be
$\frac{n}{t}$, $\frac{t}{n}$ respectively. In this way,
$\dot{S_{tot}}$ reduces to
\begin{eqnarray}\nonumber
\dot{S_{tot}}&=&\frac{8 \pi^{2} t}{n^{3}} \bigg(3 n^{2} +
U\bigg(\frac{2}{3n} - 1\bigg)\bigg) \bigg(\frac{1}{2}+
\frac{U}{2n^{2}}\bigg(\frac{2}{3n} -
1\bigg)\bigg)\\\label{37}&+&\frac{\pi t}{n}\bigg(3 +
\frac{U}{n^{2}}\bigg(\frac{2}{3n}-1\bigg)\bigg).
\end{eqnarray}
where $U=3\bigg(6\alpha(2n^{2}-n)+2\sqrt{3}\beta
n\sqrt{n^{2}-n}\bigg)$. In order to analyze the clear picture of
validity of GSLT for this entropy on the Hubble horizon, we plot
$\dot{S_{tot}}$ against cosmic time ($t$) by fixing constant
parameters as $\alpha=0.2$, $\beta=0.001$ and $n=4$ as shown in
Figure \textbf{1}. This shows that $\dot{S_{tot}}$ remains positive
with increasing value of $t$ which confirms the validity of GSLT at
apparent horizon with Bekenstein entropy.

To examine the thermodynamic equilibrium, we differentiate
Eq.(\ref{37}) to get $\ddot{S_{tot}}$ given below
\begin{eqnarray}\nonumber
\ddot{S_{tot}}&=&\frac{8 \pi^{2}}{n^{3}} \bigg(3 n^{2} +
U\bigg(\frac{2}{3n} - 1\bigg)\bigg) \bigg(\frac{1}{2}+
\frac{U}{2n^{2}}\bigg(\frac{2}{3n} - 1\bigg)\bigg)\\\label{38}
&+&\frac{\pi}{n}\bigg(3 +
\frac{U}{n^{2}}\bigg(\frac{2}{3n}-1\bigg)\bigg).
\end{eqnarray}
We plot $X=\ddot{S_{tot}}$ versus $n$ in Figure \textbf{2} which
shows that $\ddot{S_{tot}}<0$ for the selected range of $n$. Hence,
thermal equilibrium condition is satisfied for Bekenstein entropy at
apparent horizon.
\begin{figure}[h]
\begin{minipage}{14pc}
\includegraphics[width=16pc]{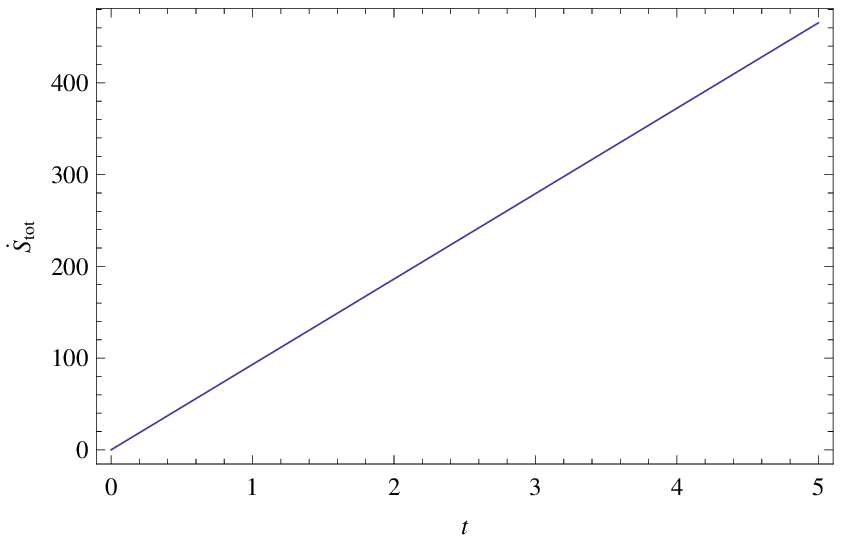}
\caption{\label{label}Plot of $\dot{S_{tot}}$ by taking Bekenstein
entropy as entropy at apparent horizon, where time is measured in
second.}
\end{minipage}\hspace{3pc}%
\begin{minipage}{14pc}
\includegraphics[width=16pc]{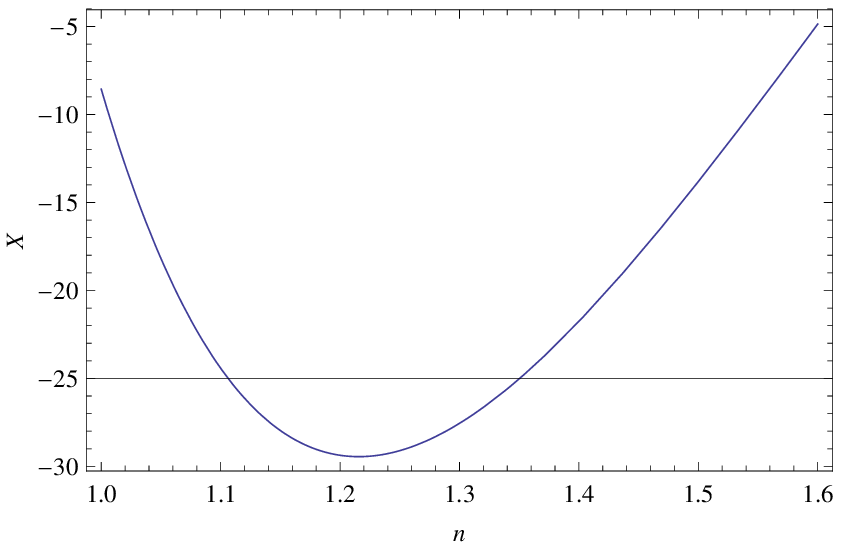}
\caption{Plot of $X=\ddot{S_{tot}}$ by taking Bekenstein entropy as
entropy at apparent horizon.}
\end{minipage}\hspace{3pc}%
\end{figure}

\subsection{Logarithmic corrections to entropy}

Logarithmic corrections arises from loop quantum gravity due to
thermal equilibrium and quantum fluctuations \cite{34}-\cite{40}.
The entropy on apparent horizon can be defined as follows
\begin{equation}\label{39}
S_{h}=\frac{A}{4G}+\eta ln\bigg[\frac{A}{4G}\bigg]-\xi
\frac{4G}{A}+\gamma,
\end{equation}
here $\eta$, $\xi$ and $\gamma$ are dimensionless constants.
Differentiating with respect to $t$, we get
\begin{equation}\label{40}
\dot{S_{h}}=\bigg(\frac{2\pi }{H}+2\eta H+\frac{2\xi H^{3}}{\pi
}\bigg)\dot{R_{h}},
\end{equation}
which takes the following form by inserting value of $\dot{R_{h}}$
from Eq.(\ref{27})
\begin{eqnarray}\nonumber
\dot{S_{h}}&=&\bigg(\frac{\pi }{H}+\eta H+\frac{\xi H^{3}}{\pi
}\bigg)\bigg(3-\bigg(\frac{1}{1-\Omega_{d}}\bigg)
\bigg(\frac{\dot{\rho_{d}}}{3H^{3}}+\frac{\rho_{d}}{H^{2}}\bigg)\\\label{41}
&&\bigg(1+\frac{\rho_{d}}{3H^{2}(1-\Omega_{d})}\bigg)^{-1}\bigg).
\end{eqnarray}

In the presence of logarithmic entropy, $\dot{S}_{tot}$ can be
obtained by using Eq.(\ref{31}) and Eq.(\ref{41})
\begin{eqnarray}\nonumber
\dot{S_{tot}}&=&\frac{8\pi^{2}}{H^{3}}\bigg(3H^{2}-\rho_{d}-\frac{\dot{\rho_{d}}}{3H}\bigg)
\bigg(\frac{1}{2}-\bigg(\frac{1}{2(1-\Omega_{d})}\bigg)\\\nonumber
&\times&\bigg(\frac{\dot{\rho_{d}}}{3H^{3}}+\frac{\rho_{d}}{H^{2}}\bigg)
\bigg(1+\frac{\rho_{d}}{3H^{2}(1-\Omega_{d})}\bigg)^{-1}\bigg)+\bigg(\frac{\pi
}{H}+\eta H+\frac{\xi H^{3}}{\pi }\bigg)\\\label{42}
&\times&\bigg(3-\bigg(\frac{1}{1-\Omega_{d}}\bigg)
\bigg(\frac{\dot{\rho_{d}}}{3H^{3}}+\frac{\rho_{d}}{H^{2}}\bigg)
\bigg(1+\frac{\rho_{d}}{3H^{2}(1-\Omega_{d})}\bigg)^{-1}\bigg).
\end{eqnarray}
By substituting value of scale factor, the above equation reduces to
\begin{eqnarray}\nonumber
\dot{S}_{tot}&=&\frac{8 \pi^{2} t}{n^{3}} \bigg(3 n^{2} +
U\bigg(\frac{2}{3n} - 1\bigg)\bigg) \bigg(\frac{1}{2}+
\frac{U}{2n^{2}}\bigg(\frac{2}{3n} - 1\bigg)\bigg)\\\label{43}
&+&\bigg(\frac{\pi t}{n}+\frac{\eta n}{t}+\frac{\xi n^{3}}{\pi
t^{3}}\bigg))\bigg(3 +
\frac{U}{n^{2}}\bigg(\frac{2}{3n}-1\bigg)\bigg).
\end{eqnarray}
Differentiating above equation, we get
\begin{eqnarray}\nonumber
\ddot{S}_{tot}&=&\frac{8 \pi^{2}}{n^{3}} \bigg(3 n^{2} +
U\bigg(\frac{2}{3n} - 1\bigg)\bigg) \bigg(\frac{1}{2}+
\frac{U}{2n^{2}}\bigg(\frac{2}{3n} - 1\bigg)\bigg)\\\label{44}
&+&\bigg(\frac{\pi}{n}-\frac{\eta n}{t^{2}}-\frac{3\xi n^{3}}{\pi
t^{4}}\bigg)\bigg(3 +
\frac{U}{n^{2}}\bigg(\frac{2}{3n}-1\bigg)\bigg).
\end{eqnarray}
\begin{figure}[h]
\begin{minipage}{14pc}
\includegraphics[width=16pc]{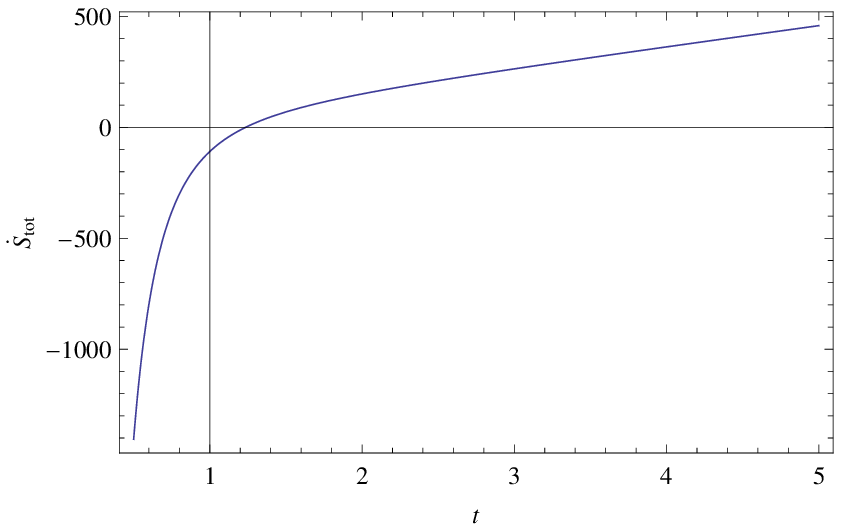}
\caption{\label{label}Plot of $\dot{S_{tot}}$ by taking Logarithmic
entropy as entropy at apparent horizon, where time is measured in
second.}
\end{minipage}\hspace{3pc}%
\begin{minipage}{14pc}
\includegraphics[width=16pc]{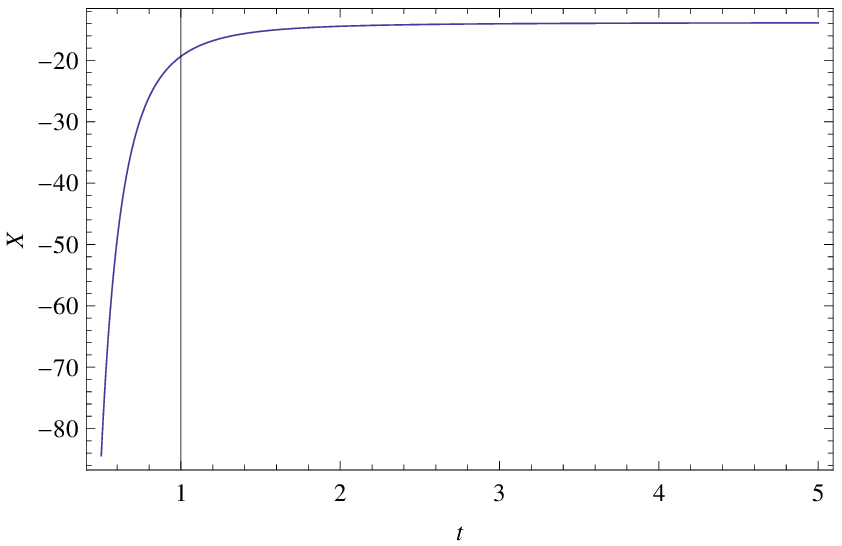}
\caption{Plot of $X=\ddot{S_{tot}}$ by taking Logarithmic entropy as
entropy at apparent horizon, where time is measured in second.}
\end{minipage}\hspace{3pc}%
\end{figure}
Figure 3 presents the plot of $\dot{S}_{tot}$ at apparent horizon by
taking logarithmic entropy at apparent horizon. Here we have taken
$\eta=3.8$ and $\xi=3$ along with same values of $\alpha$, $\beta$
and $n$ as in above mentioned case. Here $\dot{S}_{tot}$ remains
negative for $t<1.5$ while it moves in positive direction
$t\geq1.5$. Hence, validity of GSLT is verified for $t\geq1.5$ at
apparent horizon with logarithmic entropy. Figure \textbf{4} shows
that $X=\ddot{S_{tot}}<0$ with increasing value of $t$ and $n=1.5$.
Hence, for logarithmic entropy at apparent horizon, the condition of
thermal equilibrium is satisfied.

\subsection{Renyi Entropy}

A novel type of Renyi entropy was recommended by Biro and Czinner
\cite{41} on black hole horizons by considering Bekenstein-Hawking
entropy as non extensive Tsalis entropy. The modified Renyi entropy
can be defined as \cite{42}
\begin{equation}\label{45}
S_{h}=\frac{1}{\lambda}ln\bigg[1+\lambda \frac{A}{4G}\bigg].
\end{equation}
it behaves as Bekenstein entropy for $\lambda=0$. Differentiating
with respect to $t$, we obtain
\begin{equation}\label{46}
\dot{S_{h}}=\frac{2\pi H}{H^{2}+\lambda \pi}\dot{R_{h}}.
\end{equation}
Using Eq.(\ref{27}) in above equation, we get
\begin{eqnarray}\label{47}
\dot{S_{h}}=\frac{\pi H}{H^{2}+\lambda
\pi}\bigg(3-\bigg(\frac{1}{1-\Omega_{d}}\bigg)
\bigg(\frac{\dot{\rho_{d}}}{3H^{3}}+\frac{\rho_{de}}{H^{2}}\bigg)
\bigg(1+\frac{\rho_{d}}{3H^{2}(1-\Omega_{d})}\bigg)^{-1}\bigg).
\end{eqnarray}
Combining Eqs.(\ref{31})and (\ref{47}) to get
\begin{eqnarray}\nonumber
\dot{S_{tot}}&=&\frac{8(\pi)^{2}}{H^{3}}\bigg(3H^{2}-\rho_{d}-\frac{\dot{\rho_{d}}}{3H}\bigg)
\bigg(\frac{1}{2}-\bigg(\frac{1}{2(1-\Omega_{d})}\bigg)\bigg(\frac{\dot{\rho_{d}}}{3H^{3}}
+\frac{\rho_{d}}{H^{2}}\bigg)\\\nonumber
&\times&\bigg(1+\frac{\rho_{d}}{3H^{2}(1-\Omega_{d})}\bigg)^{-1}\bigg)+\frac{\pi
H}{H^{2}+\lambda \pi}\bigg(3-\bigg(\frac{1}{1-\Omega_{d}}\bigg)
\\\label{48}
&\times&\bigg(\frac{\dot{\rho_{d}}}{3H^{3}}+\frac{\rho_{d}}{H^{2}}\bigg)
\bigg(1+\frac{\rho_{d}}{3H^{2}(1-\Omega_{d})}\bigg)^{-1}\bigg).
\end{eqnarray}
For power law scale factor, we obtain
\begin{eqnarray}\nonumber
\dot{S}_{tot}&=&\frac{8 \pi^{2} t}{n^{3}} \bigg(3 n^{2} +
U\bigg(\frac{2}{3n} - 1\bigg)\bigg) \bigg(\frac{1}{2}+
\frac{U}{2n^{2}}\bigg(\frac{2}{3n} - 1\bigg)\bigg)\\\label{49}
&+&\bigg(\frac{n \pi t}{n^{2}+\pi\lambda t^{2}}\bigg)\bigg(3 +
\frac{U}{n^{2}}\bigg(\frac{2}{3n}-1\bigg)\bigg).
\end{eqnarray}
The plot of $\dot{S}_{tot}$ by taking Renyi entropy at apparent
horizon is presented by Figure \textbf{5}. Here $\alpha$, $\beta$
and $n$ has same values like previous case and $\lambda=1.5$. In
this case, $\dot{S_{tot}}$ behaves positively with the passage of
time which verifies the validity of GSLT for the present case.
Further, differentiating above equation, we get
\begin{eqnarray}\nonumber
\ddot{S_{tot}}&=&\frac{8 \pi^{2}}{n^{3}} \bigg(3 n^{2} +
U\bigg(\frac{2}{3n} - 1\bigg)\bigg) \bigg(\frac{1}{2}+
\frac{U}{2n^{2}}\bigg(\frac{2}{3n} - 1\bigg)\bigg)\\\label{50}
&+&\bigg(\frac{n^{3}\pi-\lambda \pi^{2}n t^{2}}{(n^{2}+\pi\lambda
t^{2})^{2}}\bigg)\bigg(3 +
\frac{U}{n^{2}}\bigg(\frac{2}{3n}-1\bigg)\bigg).
\end{eqnarray}
The plot of this expression is shown in Figure \textbf{6} which
shows that $\ddot{S}_{tot}<0$ for $n=1.5$ with the passage of time.
Hence, the condition for thermal equilibrium is satisfied in case of
Renyi entropy at apparent horizon.
\begin{figure}[h]
\begin{minipage}{14pc}
\includegraphics[width=16pc]{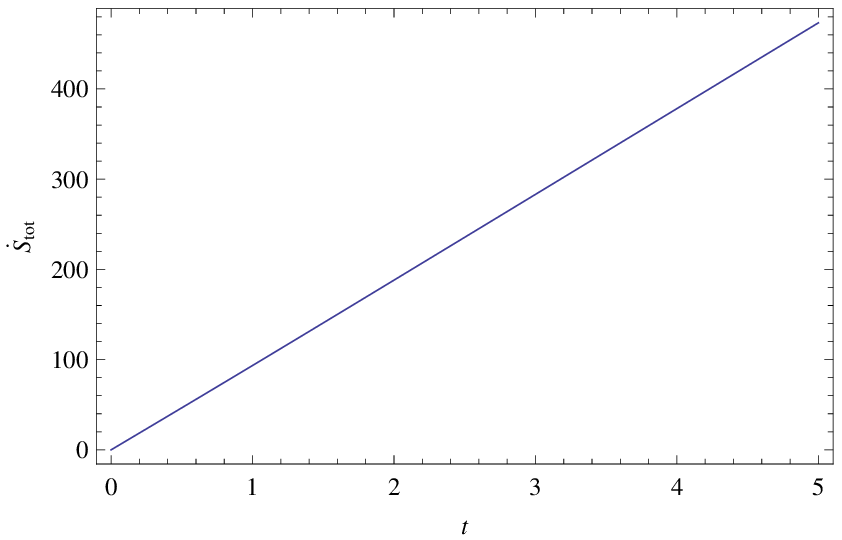}
\caption{\label{label}Plot of $\dot{S_{tot}}$ by taking Renyi
entropy as entropy at apparent horizon, where time is measured in
second.}
\end{minipage}\hspace{3pc}%
\begin{minipage}{14pc}
\includegraphics[width=16pc]{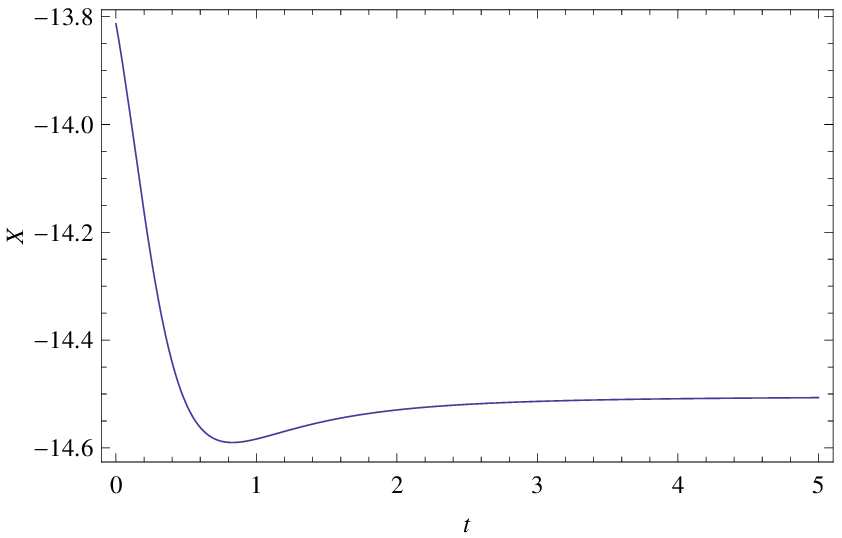}
\caption{Plot of $X=\ddot{S_{tot}}$ by taking Renyi entropy as
entropy at apparent horizon, where time is measured in second.}
\end{minipage}\hspace{3pc}%
\end{figure}

\subsection{Power Law  Entropic correction}

The power law corrections to entropy appear in dealing with
entanglement of quantum fields in and out of the horizon \cite{43}.
The corrected entropy takes the form \cite{44}
\begin{equation}\label{51}
S_{h}=\frac{A}{4G}\bigg(1-k_{\mu}A^{1-\frac{\mu}{2}}\bigg),
\end{equation}
with
$k_{\mu}=\frac{\mu(4\pi)^{\frac{\mu}{2}-1}}{(4-\mu)~r^{(2-\mu)}_{c}}$,
$r_{c}$ is crossover length and $\mu$ appears as a constant.
\begin{equation}\label{52}
\dot{S_{h}}=\frac{\pi \dot{R_{h}}
}{H}\bigg(2-k_{\mu}(4-\mu)\big(\frac{4\pi}{H^{2}}\big)^{1-\frac{\mu}{2}}\bigg).
\end{equation}
Utilization of Eq.(\ref{27}) in above equation leads to
\begin{eqnarray}\nonumber
\dot{S_{h}}&=&\frac{\pi}
{2H}\bigg(2-k_{\mu}(4-\mu)\big(\frac{4\pi}{H^{2}}\big)^{1-\frac{\mu}{2}}\bigg)
\bigg(3-\bigg(\frac{1}{1-\Omega_{d}}\bigg)
\bigg(\frac{\dot{\rho_{d}}}{3H^{3}}+\frac{\rho_{d}}{H^{2}}\bigg)\\\label{53}
&\times&\bigg(1+\frac{\rho_{d}}{3H^{2}(1-\Omega_{d})}\bigg)^{-1}\bigg).
\end{eqnarray}
Joining Eqs.(\ref{31})and (\ref{53}) to obtain
\begin{eqnarray}\nonumber
\dot{S_{tot}}&=&\frac{8(\pi)^{2}}{H^{3}}\bigg(3H^{2}-\rho_{d}-\frac{\dot{\rho_{d}}}{3H}\bigg)
\bigg(\frac{1}{2}-\bigg(\frac{1}{2(1-\Omega_{d})}\bigg)\bigg(\frac{\dot{\rho_{d}}}{3H^{3}}
+\frac{\rho_{d}}{H^{2}}\bigg)\\\nonumber
&\times&\bigg(1+\frac{\rho_{d}}{3H^{2}(1-\Omega_{d})}\bigg)^{-1}\bigg)+\frac{\pi}
{2H}\bigg(2-k_{\mu}(4-\mu)\big(\frac{4\pi}{H^{2}}\big)^{1-\frac{\mu}{2}}\bigg)\\\label{54}
&\times&\bigg(3-\bigg(\frac{1}{1-\Omega_{d}}\bigg)
\bigg(\frac{\dot{\rho_{d}}}{3H^{3}}+\frac{\rho_{d}}{H^{2}}\bigg)
\bigg(1+\frac{\rho_{d}}{3H^{2}(1-\Omega_{d})}\bigg)^{-1}\bigg).
\end{eqnarray}
In the presence of scale factor, the above expression turns out to
be
\begin{eqnarray}\nonumber
\dot{S_{tot}}&=&\frac{8 \pi^{2} t}{n^{3}} \bigg(3 n^{2} +
U\bigg(\frac{2}{3n} - 1\bigg)\bigg) \bigg(\frac{1}{2}+
\frac{U}{2n^{2}}\bigg(\frac{2}{3n} - 1\bigg)\bigg)\\\label{55} &+&
\bigg(\frac{\pi t}{2n}\bigg)\bigg(2-\frac{\mu}{(\frac{n
r_{c}}{t})^{2-\mu}}\bigg)\bigg(3
+\frac{U}{n^{2}}\bigg(\frac{2}{3n}-1\bigg)\bigg).
\end{eqnarray}
By taking power Law entropy at apparent horizon, $\dot{S_{tot}}$ is
plotted at apparent horizon as shown in Figure \textbf{7}. With same
values for $\alpha$, $\beta$ and $n$, we have taken $\mu=5$ and
$r_{c}=2$. Here the effectiveness of GSLT at apparent horizon is
certified by positive moves of $\dot{S_{tot}}$ with increasing $t$.
Differentiating above equation, we get
\begin{eqnarray}\nonumber
\ddot{S_{tot}}&=&\frac{8 \pi^{2}}{n^{3}} \bigg(3 n^{2} +
U\bigg(\frac{2}{3n} - 1\bigg)\bigg) \bigg(\frac{1}{2}+
\frac{U}{2n^{2}}\bigg(\frac{2}{3n} - 1\bigg)\bigg)\\\label{56} &+&
\bigg(\frac{\pi t}{n}-\frac{\pi \mu}{2n}(n
r_{c})^{(2-\mu)}t^{(\mu-1)}\bigg)\bigg(3
+\frac{U}{n^{2}}\bigg(\frac{2}{3n}-1\bigg)\bigg).
\end{eqnarray}
Just like above mentioned three cases, in case of power law entropy
at apparent horizon, the condition for thermal equilibrium is
satisfied with the passage of cosmic time as shown in Figure
\textbf{8}.
\begin{figure}[h]
\begin{minipage}{14pc}
\includegraphics[width=16pc]{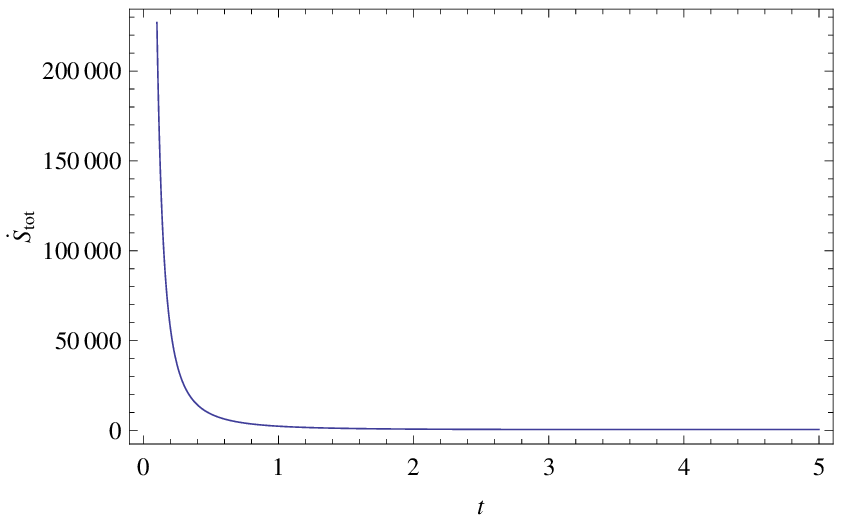}
\caption{\label{label}Plot of $\dot{S_{tot}}$ by taking Power Law
entropy as entropy at apparent horizon, where time is measured in
second.}
\end{minipage}\hspace{3pc}%
\begin{minipage}{14pc}
\includegraphics[width=16pc]{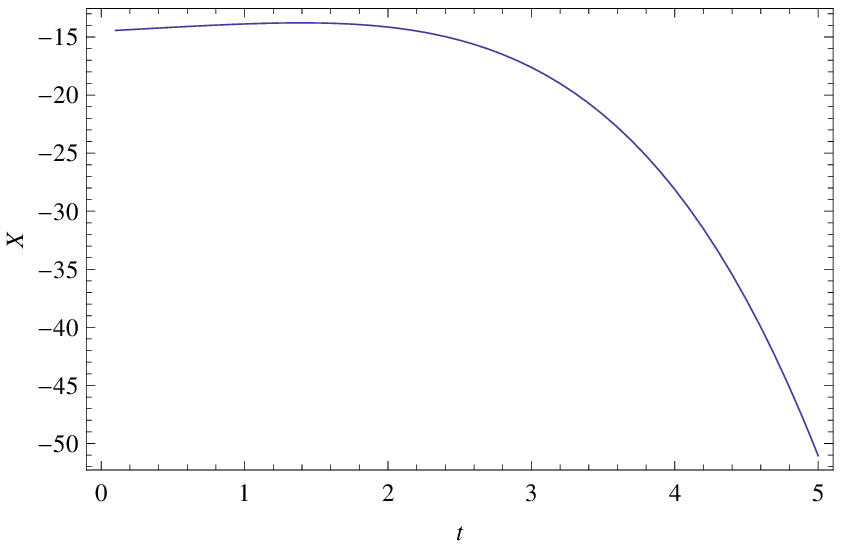}
\caption{Plot of $X=\ddot{S_{tot}}$ by taking Power Law entropy as
entropy at apparent horizon, where time is measured in second.}
\end{minipage}\hspace{3pc}%
\end{figure}

\section{Generalized Second Law of Thermodynamics at event horizon}

In this section, we study GSL of thermodynamics at event horizon
which is defined as
$R_{h}=a(t)\int^{\infty}_{t}\frac{d\hat{t}}{a(\hat{t})}$. Its
derivative with respect to time is given by $\dot{R_{h}}=H R_{h}-1$.
The temperature we used in this section is $T=\frac{bH}{2\pi}$,
where $b$ is a constant. For the present case, rewriting
Eq.(\ref{5}) by using value of $T$ and $\dot{R_{h}}$, we have
following equation for entropy inside horizon
\begin{equation}\label{57}
\dot{S_{in}}=-\frac{8\pi^{2}}{b
H}R_{h}^{2}\bigg(3H^{2}-\rho_{d}-\frac{\dot{\rho_{d}}}{3H}\bigg).
\end{equation}

\subsection{Bekenstein Entropy}

Under this scenario, Eq.(\ref{33}) can be written as
\begin{equation}\label{58}
\dot{S_{h}}=2\pi R_{h}(H R_{h}-1).
\end{equation}
The equation for $\dot{S}_{tot}$ can be obtained by using
Eqs.(\ref{53}) and (\ref{58}) as follows
\begin{equation}\label{59}
\dot{S_{tot}}=-\frac{8\pi^{2}}{bH}R_{h}^{2}\bigg(3H^{2}-\rho_{de}-\frac{\dot{\rho_{de}}}{3H}\bigg)+2\pi
R_{h}(H R_{h}-1).
\end{equation}
By putting values of scale factor and $R_{h}$ in above equation, we
have
\begin{equation}\label{60}
\dot{S_{tot}}=-\frac{8\pi^{2}t}{n(n-1)^{2}b}\bigg(3n^{2}+U\bigg(\frac{2}{3n}-1\bigg)\bigg)+2\pi
t.
\end{equation}
Differentiating above equation with respect to $t$, we get
\begin{equation}\label{61}
\dot{S_{tot}}=-\frac{8\pi^{2}}{n(n-1)^{2}b}\bigg(3n^{2}+U\bigg(\frac{2}{3n}-1\bigg)\bigg)+2\pi.
\end{equation}
\begin{figure}[h]
\begin{minipage}{14pc}
\includegraphics[width=16pc]{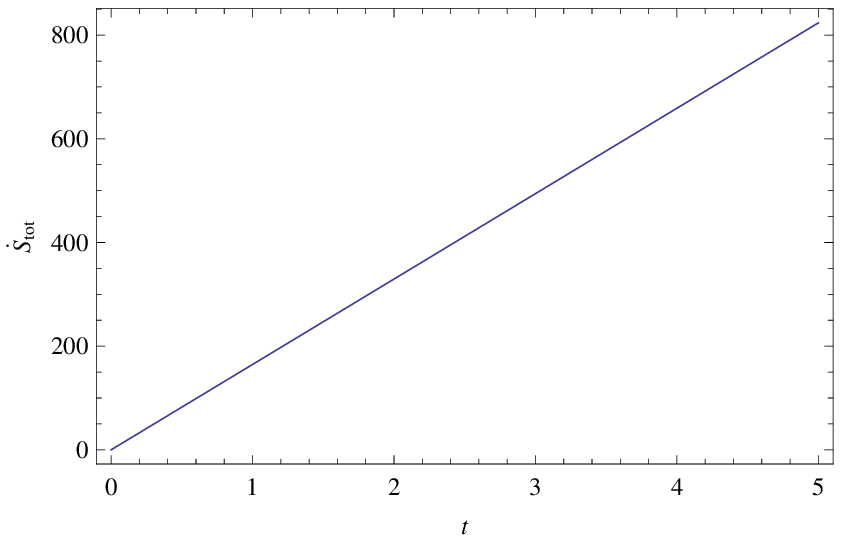}
\caption{\label{label}Plot of $\dot{S_{tot}}$ by taking Bekenstein
entropy as entropy at event horizon, where time is measured in
second.}
\end{minipage}\hspace{3pc}%
\begin{minipage}{14pc}
\includegraphics[width=16pc]{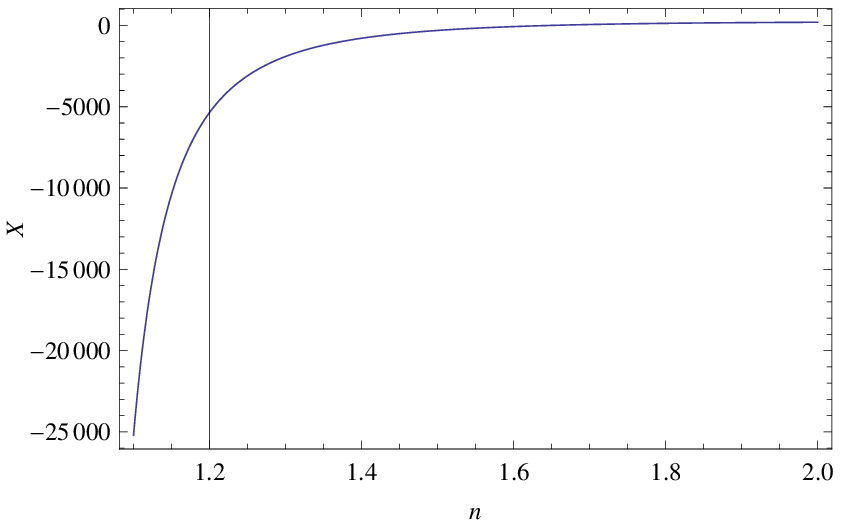}
\caption{Plot of $X=\ddot{S_{tot}}$ by taking Bekenstein entropy as
entropy at event horizon.}
\end{minipage}\hspace{3pc}%
\end{figure}

Figure \textbf{9} contains the plot of $\dot{S}_{tot}$ by taking
Bekenstein entropy at event horizon. Here we have taken
$\alpha=0.2$, $\beta=0.001$ and  $n=4$. Its clear from figure that
$\dot{S}_{tot}$ remains positive with increasing value of $t$. This
confirms the validity of GSLT at event horizon with Bekenstein
entropy. Figure \textbf{10} shows that $\ddot{S}_{tot}<0$ for
increasing values of $n$. Hence, at event horizon, the Bekenstein
entropy fulfilled the condition of thermodynamic equilibrium.

\subsection{Logarithmic Entropy}

For this entropy at event horizon, Eq.(\ref{39}) leads to
\begin{equation}\label{62}
\dot{S_{h}}=\bigg(\frac{2\pi }{H}+2\eta H+\frac{2\xi
H^{3}}{\pi}\bigg)\dot{R_{h}}.
\end{equation}
By using Eqs.(\ref{57}) and (\ref{62}), the expression of
$\dot{S}_{tot}$ can be written as
\begin{equation}\label{63}
\dot{S_{tot}}=-\frac{8\pi^{2}}{bH}R_{h}^{2}\bigg(3H^{2}-\rho_{de}-\frac{\dot{\rho_{de}}}{3H}\bigg)+\bigg(\frac{2\pi
}{H}+2\eta H+\frac{2\xi H^{3}}{\pi}\bigg)\dot{R_{h}}.
\end{equation}
The following equation is obtained by using values of scale factor
and $R_{h}$
\begin{equation}\label{64}
\dot{S_{tot}}=-\frac{8\pi^{2}t}{n(n-1)^{2}b}\bigg(3n^{2}+U\big(\frac{2}{3n}-1\big)\bigg)+\frac{2\pi
t}{(n-1)^{2}}+\frac{2\eta}{t}+\frac{2\xi(n-1)^{2}}{\pi t^{3}}.
\end{equation}
Differentiating above equation with respect to $t$, we obtain
\begin{equation}\label{65}
\ddot{S_{tot}}=-\frac{8\pi^{2}}{n(n-1)^{2}b}\bigg(3n^{2}+U\big(\frac{2}{3n}-1\big)\bigg)+\frac{2\pi
}{(n-1)^{2}}-\frac{2\eta}{t^{2}}-\frac{6\xi(n-1)^{2}}{\pi t^{4}}.
\end{equation}
\begin{figure}[h]
\begin{minipage}{14pc}
\includegraphics[width=16pc]{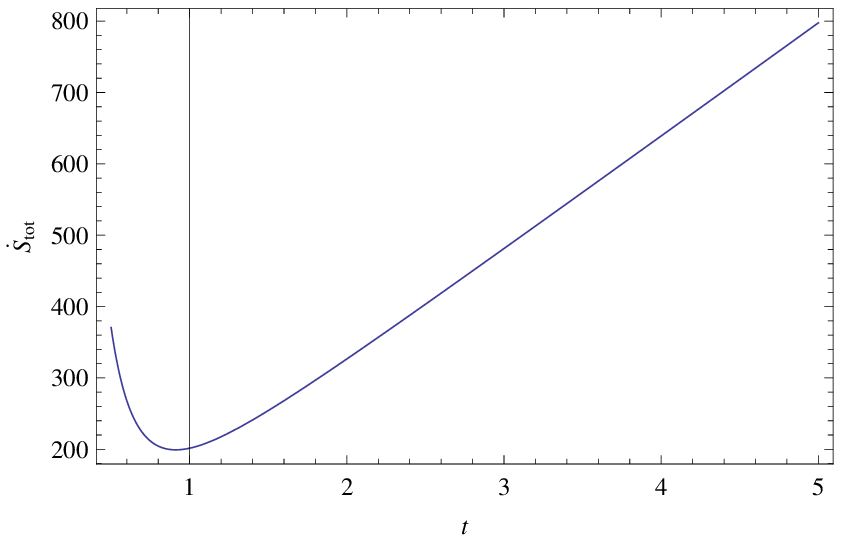}
\caption{\label{label}Plot of $\dot{S_{tot}}$ by taking Logarithmic
entropy as entropy at event horizon, where time is measured in
second.}
\end{minipage}\hspace{3pc}%
\begin{minipage}{14pc}
\includegraphics[width=16pc]{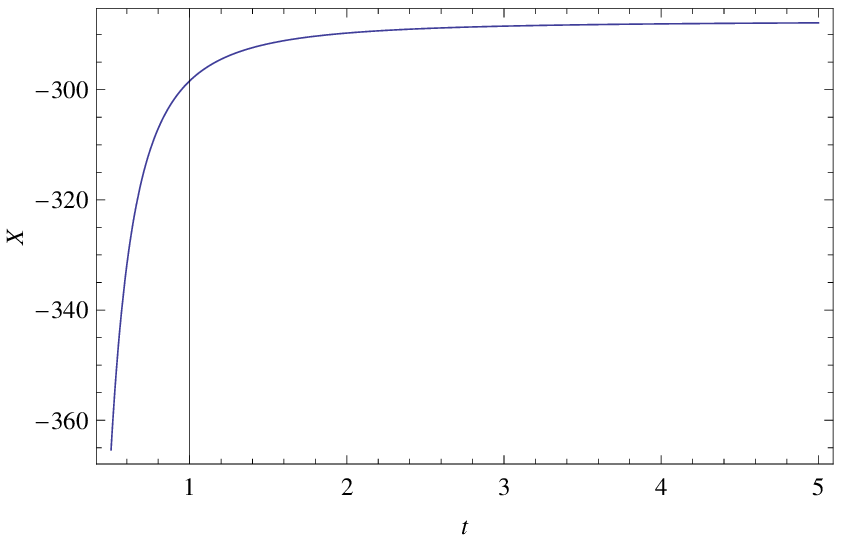}
\caption{Plot of $X=\ddot{S_{tot}}$ by taking Logarithmic entropy as
entropy at event horizon, where time is measured in second.}
\end{minipage}\hspace{3pc}%
\end{figure}

Figure \textbf{11} presents the plot of $\dot{S_{tot}}$ by taking
logarithmic entropy at event horizon. Here we have taken $\eta=4$
and $\xi=6$ along with same values of $\alpha$, $\beta$ and $n$ as
in above mentioned case. Clearly, $\dot{S_{tot}}$ moves in positive
direction as value of $t$ increases. The validity of GSLT is
verified at event horizon in the presence of logarithmic entropy.
From Figure \textbf{12}, we can see that $\ddot{S_{tot}}<0$ for
$n=1.5$. Hence, for this case, thermodynamic equilibrium condition
holds.

\subsection{Renyi Entropy}

The following form is obtained from Eq.(\ref{44}), by substituting
value for $\dot{R_{h}}$
\begin{equation}\label{66}
\dot{S_{h}}=\frac{2\pi H}{H^{2}+\lambda \pi}(H R_{h}-1).
\end{equation}
Joining Eqs.(\ref{57}) and (\ref{66}), we get
\begin{equation}\label{67}
\dot{S_{tot}}=-\frac{8\pi^{2}}{bH}R_{h}^{2}\bigg(3H^{2}-\rho_{de}-\frac{\dot{\rho_{de}}}{3H}\bigg)+\frac{2\pi
H}{H^{2}+\lambda \pi}(H R_{h}-1).
\end{equation}
By using values of scale factor and $R_{h}$, above equation reduces
to
\begin{equation}\label{68}
\dot{S_{tot}}=-\frac{8\pi^{2}t}{n(n-1)^{2}b}\bigg(3n^{2}+U\big(\frac{2}{3n}-1\big)\bigg)+\frac{2\pi
t}{(n-1)^{2}+\lambda \pi t^{2}}.
\end{equation}
Differentiating above equation with respect to $t$, we get
\begin{equation}\label{69}
\ddot{S_{tot}}=-\frac{8\pi^{2}}{n(n-1)^{2}b}\bigg(3n^{2}+U\big(\frac{2}{3n}
-1\big)\bigg)+\frac{2\pi(n-1)^{2}-2\pi^{2}\lambda
t^{2} }{((n-1)^{2}+\lambda \pi t^{2})^{2}}.
\end{equation}
\begin{figure}[h]
\begin{minipage}{14pc}
\includegraphics[width=16pc]{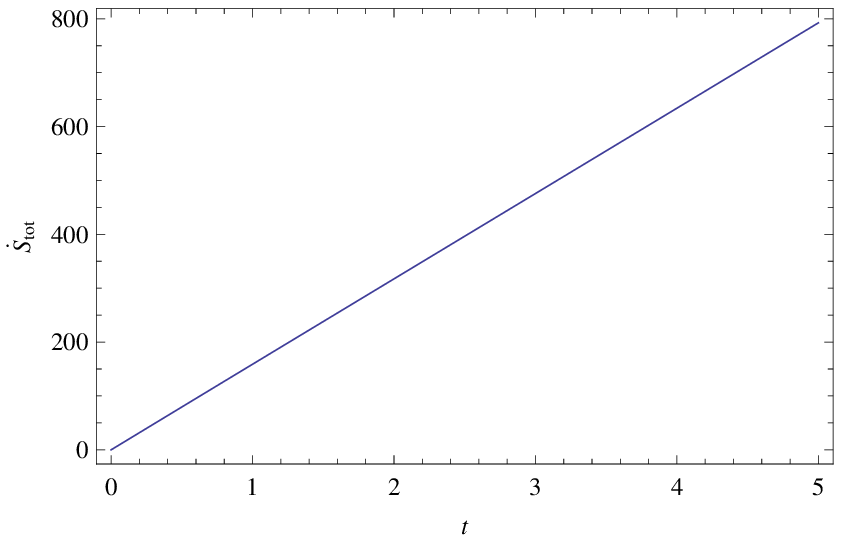}
\caption{\label{label}Plot of $\dot{S_{tot}}$ by taking Renyi
entropy as entropy at event horizon.}
\end{minipage}\hspace{3pc}%
\begin{minipage}{14pc}
\includegraphics[width=16pc]{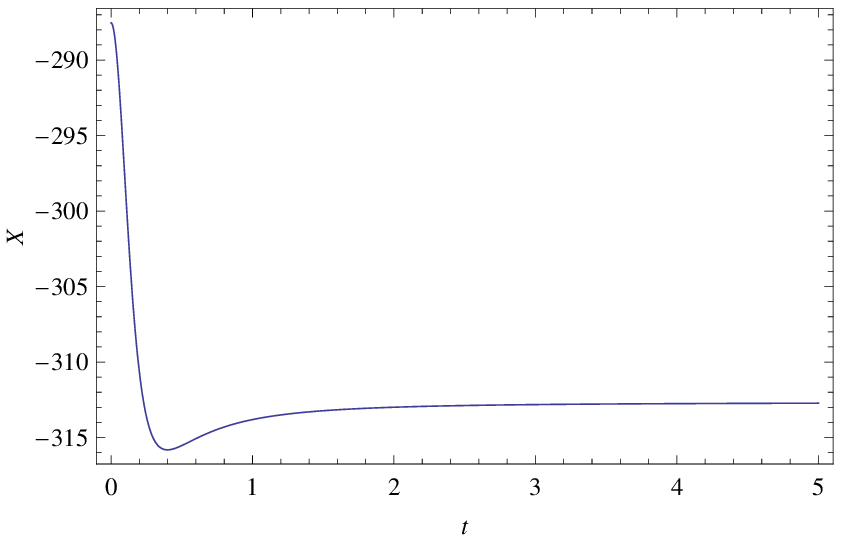}
\caption{Plot of $X=\ddot{S_{tot}}$ by taking Renyi entropy as
entropy at event horizon.}
\end{minipage}\hspace{3pc}%
\end{figure}
The plot of $\dot{S_{tot}}$ for Renyi entropy at event horizon is
presented in Figure \textbf{13}. Here $\alpha$, $\beta$ and $n$ has
same values like previous case while $\lambda=1.5$. In this case,
$\dot{S_{tot}}$ behaves positively with the passage of time which
verifies the validity of GSLT. Figure \textbf{14} shows that the
trajectories of $\ddot{S_{tot}}$ remains negative for increasing of
$t$ with $n=1.5$. This means that the present scenario obeys the
condition for thermodynamic equilibrium.

\subsection{Power law  Entropy}

Under conditions of present section, Eq.(\ref{51}) reduces to
\begin{equation}\label{70}
\dot{S_{h}}=\frac{\pi}
{H}\bigg(2-k_{\mu}(4-\mu)\big(\frac{4\pi}{H^{2}}\big)^{1-\frac{\mu}{2}}\bigg)(H
R_{h}-1).
\end{equation}
Joining Eqs.(\ref{57}) and (\ref{70}) to get the following equation
\begin{eqnarray}\label{71}
\dot{S_{tot}}=-\frac{8\pi^{2}}{bH}R_{h}^{2}\bigg(3H^{2}-\rho_{de}-\frac{\dot{\rho_{de}}}{3H}\bigg)+\frac{\pi}
{H}\bigg(2-k_{\mu}(4-\mu)\big(\frac{4\pi}{H^{2}}\big)^{1-\frac{\mu}{2}}\bigg)(H
R_{h}-1).
\end{eqnarray}
Inserting conditions for scale factor and $R_{h}$ in above equation,
we get
\begin{equation}\label{72}
\dot{S_{tot}}=-\frac{8\pi^{2}t}{n(n-1)^{2}b}\bigg(3n^{2}+U\big(\frac{2}
{3n}-1\big)\bigg)+\bigg(2-\mu\bigg(\frac{t}{r_{c}(n-1)}\bigg)^{2-\mu}\bigg)
\frac{\pi t}{(n-1)^{2}}.
\end{equation}
The plot of this expression is displayed in Figure \textbf{15} with
same values for $\alpha$, $\beta$ and $n$ while $\mu=5$ and
$r_{c}=2$. Here the effectiveness of GSLT at event horizon is
certified by positive moves of $\dot{S_{tot}}$ with increasing $t$.
Differentiating with respect to $t$, we obtain
\begin{equation}\label{73}
\ddot{S_{tot}}=-\frac{8\pi^{2}}{n(n-1)^{2}b}\bigg(3n^{2}+U\big(\frac{2}{3n}-1\big)\bigg)+\frac{2\pi}{(n-1)^{2}}-
\frac{\mu \pi(3-\mu)t^{2-\mu}}{(r_{c}(n-1))^{2-\mu}(n-1)^{2}}.
\end{equation}
Figure \textbf{16} shows that the present scenario fulfils the
thermodynamic equilibrium condition for power law entropy at event
horizon.
\begin{figure}[h]
\begin{minipage}{14pc}
\includegraphics[width=16pc]{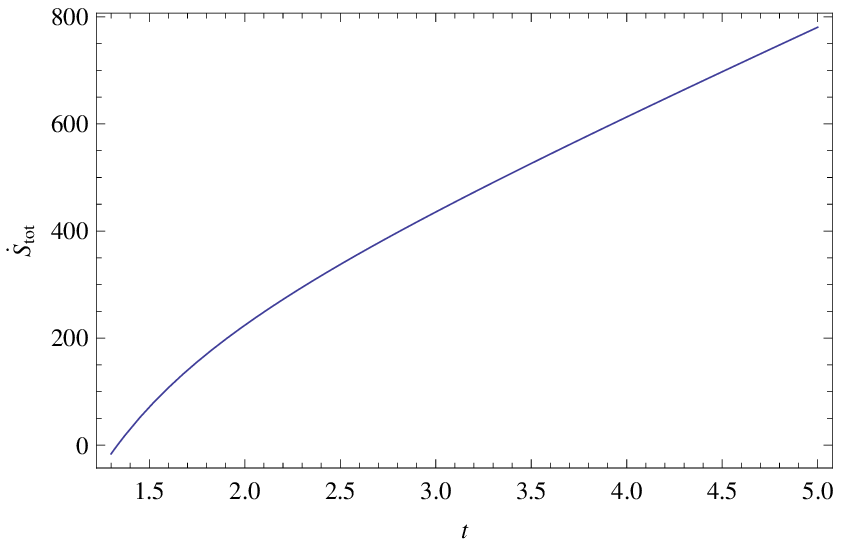}
\caption{\label{label}Plot of $\dot{S_{tot}}$ by taking Power Law
entropy as entropy at event horizon.}
\end{minipage}\hspace{3pc}%
\begin{minipage}{14pc}
\includegraphics[width=16pc]{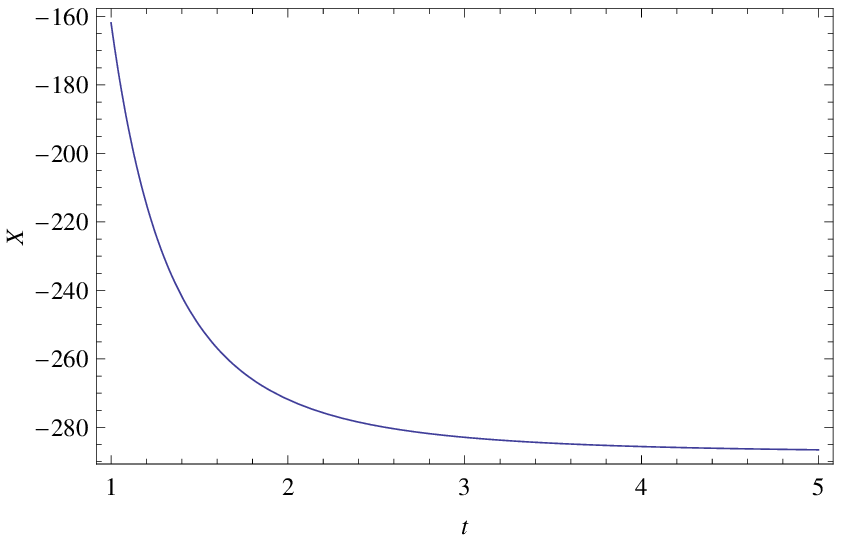}
\caption{Plot of $X=\ddot{S_{tot}}$ by taking Power Law entropy as
entropy at event horizon.}
\end{minipage}\hspace{3pc}%
\end{figure}

\section{Conclusion}

The concept of thermodynamics in cosmological system originates
through black hole physics. It was suggested \cite{10J} that the
temperature of Hawking radiations emitting from black holes is
proportional to their corresponding surface gravity on the event
horizon. Jacobson \cite{11J} found a relation between thermodynamics
and the Einstein field equations. He derived this relation on the
basis of entropy-horizon area proportionality relation along with
first law of thermodynamics (also called Clausius relation)
$dQ=TdS$, where $dQ,~T$ and $dS$ indicate the exchange in energy,
temperature and entropy change for a given system. It was shown that
the field equations for any spherically symmetric spacetime can be
expressed as $TdS=dE+PdV$ ($E,~P$ and $V$ represent the internal
energy, pressure and volume of the spherical system) for any horizon
\cite{12J}. By utilizing this relation, GSLT has been studied
extensively in the scenario of expanding behavior of the universe.
In order to discuss GSLT, horizon entropy of the universe can be
taken as one quarter of its horizon area \cite{14J} or power law
corrected \cite{15J} or logarithmic corrected \cite{16J} forms. Many
people have explored the validity of GSLT of different systems
including interaction of two  fluid components like DE and dark
matter \cite{17J}, as well as interaction of  three components of
fluid \cite{18J} in the FRW universe by using simple horizon entropy
of the universe. The thermodynamical analysis widely performed in
modified theories of gravity \cite{sj}.

Motivated by above mentioned works, we have considered a newly
proposed DE model named as Ricci Gauss-Bonnet DE in flat FRW
universe. We have developed thermodynamical quantities and analyzed
the validity of GSLT and thermodynamic equilibrium. For dense
elaboration of thermodynamics of present DE model, we have assumed
various entropy corrections such as Bekenstein entropy, Logarithmic
corrected entropy, Renyi entropy and power law entropy at apparent
horizon as well as event horizon of the universe. We have found that
GSLT holds for all cases of entropies as well as horizons. Also,
thermal equilibrium condition satisfied under certain conditions on
constant parameters. The detailed of results are as follows:

\subsection*{\underline{On Apparent Horizon}}

By utilizing usual entropy, GSLT on the apparent horizon has shown
in Figure \textbf{1} which shows that $\dot{S_{tot}}$ remains
positive with increasing value of $t$ and confirms its validity.
Figure \textbf{2} has also indicated that thermal equilibrium
condition is satisfied for Bekenstein entropy at apparent horizon.
For logarithmic corrected entropy, GSLT on apparent horizon has
displayed in Figure \textbf{3} which exhibits that GSLT remains
valid for $t\geq1.5$. However, Figure \textbf{4} shows that
$X=\ddot{S_{tot}}<0$ with increasing value of $t$ and $n=1.5$.
Hence, for logarithmic entropy at apparent horizon, the condition of
thermal equilibrium is satisfied.

The plot of $\dot{S}_{tot}$ by taking Renyi entropy at apparent
horizon has displayed in Figure \textbf{5} which behaves positively
with the passage of time and exhibits the validity of GSLT. Also,
for this entropy, the condition for thermal equilibrium has been
satisfied in case of Renyi entropy at apparent horizon (Figure
\textbf{6}). By taking power Law entropy at apparent horizon,
$\dot{S_{tot}}$ is plotted at apparent horizon as shown in Figure
\textbf{7}. Here the effectiveness of GSLT at apparent horizon is
certified by positive moves of $\dot{S_{tot}}$ with increasing $t$.
Just like above mentioned three cases, in case of power law entropy
at apparent horizon, the condition for thermal equilibrium is
satisfied with the passage of cosmic time as shown in Figure
\textbf{8}.

\subsection*{\underline{On Event Horizon}}

It has been observed from Figure \textbf{9} that GSLT remains valid
at event horizon with Bekenstein entropy. Also, at event horizon,
the Bekenstein entropy fulfilled the condition of thermodynamic
equilibrium (Figure \textbf{10}). The validity of GSLT is verified
at event horizon in the presence of logarithmic entropy (Figure
\textbf{11}). From Figure \textbf{12}, we can see that
$\ddot{S_{tot}}<0$ for $n=1.5$ which leads to the validity of
thermal equilibrium condition.

The plot of $\dot{S_{tot}}$ for Renyi entropy at event horizon is
presented in Figure \textbf{13}. It is observed that $\dot{S_{tot}}$
behaves positively with the passage of time which verifies the
validity of GSLT. Figure \textbf{14} shows that the trajectories of
$\ddot{S_{tot}}$ remains negative for increasing of $t$ with
$n=1.5$. This means that the present scenario obeys the condition
for thermodynamic equilibrium. The plot of $\dot{S_{tot}}$ for power
law corrected entropy is displayed in Figure \textbf{15} and observe
that GSLT holds in this case. Figure \textbf{16} shows that the
present scenario fulfils the thermodynamic equilibrium condition for
power law entropy at event
horizon.\\

\textbf{The authors declare that there is no conflict of interest
regarding the publication of this paper.}

\end{document}